\documentclass{elsart}
\usepackage{graphicx}
\usepackage{amsmath}
\usepackage{amsfonts}
\usepackage{amssymb}
\begin{document}
\begin{frontmatter}
\title
{A note on Tachyon dynamics}
\author{F. Canfora}
\address
{Istituto Nazionale di Fisica Nucleare, Sezione di Napoli, GC di Salerno\\
Dipartimento di Fisica ''E.R.Caianiello'', Universit\`{a} di
Salerno\\ Via S.Allende, 84081 Baronissi (Salerno), Italy\\
e-mail: canfora@sa.infn.it}
\begin{abstract}
A suitable splitting of tachyon field equation is able to disclose
non trivial properties of Born-Infeld (some known, some
unexpected) and Polyakov actions;
the tachyon equation also can be analyzed in some detail. The analysis
displays an intriguing connection between sine-Gordon theory and some crucial issues such
as the emergence of perturbative string states when a D-Brane and an anti-D-Brane
annihilate and the confinement of charged D-Branes.
\end{abstract}
\end{frontmatter}

\section{Introduction}

\noindent Tachyon dynamics is an attention-getting topic for (at least) three
reasons: firstly, it has been argued (see, in particular, \cite{Se02I} and
references therein)\ that a deeper understanding of tachyon dynamics and of
how the tachyon ''rolls down''\ from the local maxima (related to unstable
Branes) could shed more light on the vacuum of string field theory. In second
place, tachyon dynamics could also have highly non trivial consequences both
in early time (in connection with the \textit{inflationary scenario}%
\footnote{Some problems of a would be ''tachyonic''\ inflation (such as the
problem of high density perturbations and reheating) pointed out in
\cite{KL02} can be accounted for as described, for example, in \cite{KKLT03}
\cite{GST04}.}) and in late time (in connection with viable models of
\textit{dark energy} and \textit{dark matter}) cosmology (see, for an
hopefully incomplete list of papers, \cite{AL04} and references therein).
Moreover, as it has been shown, for example, in \cite{BR00} \cite{Ga00}
\cite{Klu00} \cite{GHY00} \cite{KN03}, tachyon dynamics is very well described
by Born-Infeld type actions which are very interesting in themselves.

Here it is shown that a suitable splitting of the tachyon equation of motion
into two equations (which, physically, play different roles in determining the
tachyon evolution) on the one hand allows, in many cases, a non approximate
treatment. On the other hand, it reveals an interesting and fruitful relation
between sine-Gordon theory, tachyon dynamics and several conjectures related
to tachyon condensation such as the equivalence between a tachyonic soliton
(kink) on the world-volume of an unstable D(p+1)-Brane and a BPS Dp-Brane, the
fact that, under certain circumstances, when a D-Brane and an anti-D-Brane
annihilate perturbative string states can emerge.

The paper is organized as follows: in the second section some features of
tachyon dynamics are pointed out and a suitable splitting of the equation of
motion into two equations is analyzed. In the third section some examples are
shown. In the fourth section, the relations between the above splitting,
integrable models and several non perturbative features of D-Branes dynamics
are displayed.

\section{Some Features of Tachyon Dynamics}

The tachyon dynamics is very well described (see, for example, \cite{BR00}
\cite{Ga00} \cite{Klu00} \cite{GHY00} \cite{KN03}) by the following
Born-Infeld type action:
\begin{align}
S  &  =\frac{\kappa}{g_{S}}\int\sqrt{-g}d^{D}xV(T)\sqrt{1+\alpha T^{c}T_{c}%
}\label{a1}\\
T^{c}  &  =\left(  \nabla T\right)  ^{c}=\nabla^{c}T,\quad T_{c}=\left(
\nabla T\right)  _{c}=\partial_{c}T\nonumber
\end{align}
from which one can deduce the equations of motion
\begin{equation}
\nabla_{a}T^{a}-\frac{T^{b}T^{d}}{1+T^{c}T_{c}}\nabla_{\left(  b\right.
}T_{\left.  d\right)  }=\frac{V_{,T}}{V} \label{a2}%
\end{equation}
where $g$ is the metric, $\kappa$ is a dimensional constant which depends on
the details of the string theory one is considering, $g_{S}$ is the string
coupling constant, $\sqrt{-g}d^{D}x$ is the volume element, $\nabla^{c}$ is
the Levi-Civita connection, $V(T)$ is the tachyon potential, the round
brackets denote symmetrizations and, in this section, the string length
$\alpha^{1/2}$ and $g_{S}$ will be set equal to one. If the left hand side of
eq. (\ref{a2}) vanishes then a purely electric Born-Infeld theory is
recovered\footnote{Indeed, when the tachyon potential is constant, the above
equation is able to describe more general BI dynamics with both electric and
magnetic field with the electric component orthogonal to the magnetic one
(see, for example, \cite{Gi97}).}. It is useful to rewrite eq. (\ref{a2}) in
this way
\begin{equation}
\nabla_{a}T^{a}-\frac{T^{b}T^{d}}{1+T^{c}T_{c}}\left(  L_{\nabla T}g\right)
_{bd}=\frac{V_{,T}}{V} \label{a3}%
\end{equation}
where $L_{\nabla T}g$ is the Lie derivative of the metric along the gradient
$\nabla T$ of the tachyon field $T$. From eq. (\ref{a2}) it is clear that the
second term on the left hand side of eq. (\ref{a3}), which is responsible for
caustics formation \cite{FK02}, vanishes if $\nabla T$ is a Killing vector
field of $g$ (but this is a too restrictive condition on $T$); more generally,
it vanishes if $\nabla T$ has no components along $L_{\nabla T}g$ (this
happens, for example, when $\nabla T$ is a null conformal Killing field, this
case is interesting from the cosmological point of view): when this is the
case, eq. (\ref{a3}) becomes an almost linear partial differential equation
(PDE henceforth). The following necessary condition for the absence caustics
is rich of physical consequences
\begin{equation}
\ \frac{T^{b}T^{d}}{1+T^{c}T_{c}}\left(  L_{\nabla T}g\right)  _{bd}%
=0\ \wedge\ L_{\nabla T}g\neq0. \label{nocau}%
\end{equation}
Important examples related both to Born-Infeld theory and tachyon dynamics in
which eq. (\ref{nocau}) holds will be described below. Here we want to stress
that when eq. (\ref{nocau}) is fulfilled, eq. (\ref{a3}) becomes
\begin{align}
\nabla_{a}\nabla^{a}T  &  =W_{,T},\label{eff1}\\
\ W  &  =\ln(\left|  V(T)\right|  )\nonumber
\end{align}
and this equation can be deduced from the action
\begin{equation}
S=\int\sqrt{-g}d^{D}x\left(  T^{a}T_{a}-W\right)  . \label{aziobps}%
\end{equation}
which, under rather general conditions on the tachyon potential, admits
kink-like solutions\footnote{Tachyonic solitons have been investigated, for
example, in \cite{Mu02}.}. Thus, it is fruitful to split eq. (\ref{a2}) into
two equations: eq. (\ref{eff1}) and eq. (\ref{nocau}). Of course, there are
solutions of eq. (\ref{a2}) which are not solutions of eq. (\ref{eff1}) and
eq. (\ref{nocau}) separately. Nevertheless, this splitting allows to derive
interesting properties of Born-Infeld theory in a simple way and, mainly, it
sheds more light on the interplay between perturbative and non perturbative
physics in string and Branes theory.

\section{Explicit examples}

Here, some examples in which the above splitting turns out to be useful will
be displayed. They will clarify that, in interesting cases, the role of eq.
(\ref{nocau}) is to single out solutions of eq. (\ref{eff1}) fulfilling
natural boundary conditions. Thus, when dealing with tachyon dynamics, eq.
(\ref{eff1}) and eq. (\ref{nocau}) cannot be treated on an equal footing: one
should think at eq. (\ref{eff1}) as the truly dynamical equation while eq.
(\ref{nocau}) acts as a sort of constraint.

\subsection{The Born-Infeld case}

When in eq. (\ref{eff1}) the tachyon potential is constant, one is dealing
with a Born-Infeld dynamics in which the electric field is orthogonal to the
magnetic field \cite{Gi97}. For the sake of clarity, in this subsection a flat
background metric will be considered. On a flat background, eq. (\ref{nocau})
implies%
\begin{align}
\left(  T_{t}\right)  ^{2}T_{tt}+\sum\left(  T_{j}\right)  ^{2}T_{jj} &
=2\left(  -\underset{k<j}{\sum}T_{k}T_{j}T_{kj}+\underset{j}{\sum}T_{j}%
T_{t}T_{jt}\right)  \label{bi0}\\
ds^{2} &  =-\sum\left(  dx^{j}\right)  ^{2}+\left(  dt\right)  ^{2}\nonumber
\end{align}
Let us search for the moment an elementary solution of eq. (\ref{eff1}) with
$V(T)=const$ and eq. (\ref{bi0}) of the form
\begin{equation}
T(x,t)=a_{(1)}\sin(k_{(1)}x-\omega_{(1)}t)+a_{(2)}\cos(k_{(2)}x-\omega
_{(2)}t).\label{bi3}%
\end{equation}
At a first glance, this case is non trivial because eq. (\ref{bi0}) could give
rise to restrictions on the relative amplitudes. From eq. (\ref{eff1}) with
$V(T)=const$ one gets%
\[
k_{(i)}=\pm\omega_{(i)},\ i=1,2.
\]
In fact, when one plugs eq. (\ref{bi3}) into eq. (\ref{bi0}), eq. (\ref{bi0})
is identically fulfilled with no restrictions on the amplitudes. This is a
well known phenomenon in this sector of Born-Infeld theory: when two beams
''interact'', the two beams pass through one another with at most a time delay
(see, for example, \cite{Gi97} and references therein); it is worth to note
the simplicity of the present derivation. It is also trivial to see that any
function of the form%
\begin{equation}
f=f(\overrightarrow{k}\overrightarrow{x}-\omega t)\label{big1}%
\end{equation}
solves both eq. (\ref{eff1}) with $V(T)=const$ and eq. (\ref{bi0}) if and only
if%
\begin{equation}
\left(  \overrightarrow{k}\right)  ^{2}=\omega^{2}.\label{big2}%
\end{equation}
One could think that, in this sector, Born-Infeld theory is rather similar to
Maxwell theory: in \cite{Gi97} it has been argued that the property that any
function of the form (\ref{big1}) fulfilling (\ref{big2}) solves both eq.
(\ref{eff1}) with $V(T)=const$ and eq. (\ref{bi0}) is related to the
possibility to rewrite the Born-Infeld action in this sector as a Nambu-Goto
action in which the modes decouple. Rather, it can be shown with a simple
example that the two theories, Born-Infeld and Maxwell, can be noticeably
distinguished in this sector too. Let us consider a function of the form (here
the four dimensional case will be considered but the following results hold in
any dimension)
\begin{align}
F &  =f(z-t)h(x,y),\quad\left(  \partial_{x}^{2}+\partial_{y}^{2}\right)
h=0\Rightarrow\label{big3}\\
\nabla_{a}\nabla^{a}F &  =\left[  \left(  \partial_{x}^{2}+\partial_{y}%
^{2}\right)  +\left(  \partial_{z}^{2}-\partial_{t}^{2}\right)  \right]
F=0.\nonumber
\end{align}
This is not an academic exercise since functions of this form, besides to be
completely lawful solutions of Maxwell equations, arise, for example, in the
theory of PP-waves which has been widely analyzed in connection with string
theory. From the previous discussion, one could naively think that $F$ in eq.
(\ref{big3}) also solves eq. (\ref{bi0}). If $F$ in eq. (\ref{big3}) is
inserted in eq. (\ref{bi0}) the resulting condition on $F$ is%
\begin{equation}
\left(  h_{x}\right)  ^{2}h_{xx}+\left(  h_{y}\right)  ^{2}h_{yy}+2h_{x}%
h_{y}h_{xy}=0\label{big4}%
\end{equation}
and the above condition is in general not fulfilled by non constant harmonic
functions. At a first glance, this result seems to be not consistent with some
already quoted known results \cite{Gi97}, in fact, there is no contradiction.
The properties of ''modes decoupling''\ analyzed in that reference cannot be
applied to a function $F$ of the form in eq. (\ref{big3}) since such a $F$ is
not \textit{square integrable} (due to the presence of the harmonic function
$h$) and, therefore, cannot be expanded in \textit{Fourier} series: the
concept of ''\textit{modes}''\ is not well defined for such a $F$.
Consequently, functions of the above form solve eq. (\ref{eff1}) with
$V(T)=const$ but do not fulfil the Born-Infeld equations of motion; because of
the presence of a harmonic function in the spatial coordinates $(x,y,..)$
transverse to the propagation direction $z$, such a $F$ cannot approach, for
fixed $z$ and $t$, to a constant when $x$ and $y$ are large (a harmonic
function which approaches to a constant is constant everywhere). One could
allow some $\delta-$like singularities in the $x-y$ plane since singular
harmonic functions approaching to constants exist but, again, they diverge at
the singularities. Indeed, in Born-Infeld theory the electric field has a
maximum value so that unbounded solutions are not allowed. One could think to
smooth out the $\delta-$like singularities with suitable sources and to
represent $F$ as in eq. (\ref{big3}) only in the asymptotic region far away
the singularities where $F$ is bounded. As it is clear form eq. (\ref{big4}),
in any case Born-Infeld theory does not allow $F$ of the form in eq.
(\ref{big3}). In this sector of Born-Infeld theory factorized solutions of the
form in eq. (\ref{big3}) are not allowed, and the role of eq. (\ref{bi0}) is
to ''detect'' these unwanted\ solutions of eq. (\ref{eff1}) (which, on the
basis of a naive interpretation of the results in \cite{Gi97}, could appear
perfectly lawful solutions of Born-Infeld theory). Such a dynamical mechanism
could also be related to the results \cite{BH93II} \cite{Ho94}: eq.
(\ref{bi0}) can be connected to the extrinsic curvature of a suitable
hypersurface whose ''minimality''\ is equivalent to the Born-Infeld equations.
The above ''factorized''\ solutions could not give rise to minimal
hypersurfaces. One could think that the role of eq. (\ref{bi0}) manifests
itself when the Born-Infeld field does depend, at least, on four coordinates
in such a way to allow two dimensional manifolds transverse to the propagation direction.

In fact, eq. (\ref{bi0}) is already at work in the ordinary Polyakov actions
for strings in flat space-times (which, at a first glance, has nothing to do
with the tachyon or the Born-Infeld actions in which eq. (\ref{bi0}) plays a
role) even if it does not appear explicitly into the field equations. The
Polyakov action for free strings in $D-$dimensional flat space-times looks
like the sum of $D-2$ free scalar fields actions
\begin{align}
S_{P} &  \sim\int d\tau d\sigma\underset{i}{\overset{D-2}{\sum}}\left[
\left(  \partial_{\tau}X^{i}\right)  ^{2}-\left(  \partial_{\sigma}%
X^{i}\right)  ^{2}\right]  ,\ \delta S_{P}=0\Rightarrow\label{polact}\\
0 &  =\left(  \partial_{\tau}^{2}-\partial_{\sigma}^{2}\right)  X^{i}%
\label{2dim}%
\end{align}
where, as usual, $\sigma$ and $\tau$ are the world-sheet coordinates and the
classical equations of motion for the string modes are nothing but two
dimensional wave equation. The standard quantization procedure is based on the
modes expansion of the solutions of eq. (\ref{2dim}): one promotes the
\textit{Fourier} coefficients of the expansion to creations-annihilations
operators after that the well known machinery follows. On the other hand, one
can consider solutions of eq. (\ref{2dim}) of ''unusual''\ form such as%
\begin{equation}
X^{i}(\tau,\sigma)\sim\tau^{2}+\sigma^{2},\ \tau^{3}+3\tau\sigma
^{2}\ \ and\ so\ on.\label{2dim1}%
\end{equation}
Even if the above functions have ''nothing wrong''\ as solutions of eq.
(\ref{2dim}), it is nevertheless clear why one should discard them: functions
of the form (\ref{2dim1}) cannot fulfil the standard boundary conditions. The
previous discussion suggests a different point of view: functions of the form
(\ref{2dim1}) cannot be considered because they do not fulfil eq. (\ref{bi0}).
Thus, eq. (\ref{bi0}) is able to select precisely the solutions of the
equations of motion which can fulfil the physical boundary conditions. The
fact that eq. (\ref{bi0}) plays a ''hidden'' role also for the action
(\ref{polact}) should, in fact, not be too surprising: the Polyakov action can
be obtained, \textit{via Lagrange} multipliers, from the Nambu-Goto action%
\[
S_{N-G}\sim\int d\tau d\sigma\sqrt{\det h_{ab}}%
\]
which describes the embedding of minimal two-dimensional surfaces in $D$
dimensional (flat in this case) space-times, $h_{ab}$ being the induced metric
on the surface and, indeed, the Nambu-Goto action is of Born-Infeld type for
which eq. (\ref{bi0}) plays a role. Eq. (\ref{bi0}) (eq. (\ref{nocau}) on
curved space-times) simply tells that the solutions of eq. (\ref{2dim}) are
solutions of the Nambu-Goto equations only when they fulfil natural
(\textit{Neumann} or \textit{Dirichlet}) boundary conditions.

\subsection{The tachyon case}

When the background metric is flat and the tachyon potential is non trivial,
there are no common solutions of eq. (\ref{eff1}) and eq. (\ref{bi0}) of the
form%
\begin{equation}
T=T(z-v_{0}t),\ v_{0}=const \label{sinsol}%
\end{equation}
where $v_{0}$ would be the soliton velocity. The reason is that when one plugs
the above function into eq. (\ref{bi0}) it follows that $v_{0}=1$ and in this
case eq. (\ref{sinsol}) cannot solve eq. (\ref{eff1}) with a non vanishing
tachyon potential. In fact, common solutions of eq. (\ref{eff1}) and eq.
(\ref{bi0}) can be found: to see this, let us search for solutions of the form%
\begin{align}
T  &  =T(z-v(t)),\quad a=\overset{.}{v}=\partial_{t}v,\ \label{susol1}\\
T^{\prime\prime}(z)  &  =W(T(z)) \label{susol2}%
\end{align}
which can be understood as deformations of a single soliton solution
(\ref{susol2}). If one plugs eq. (\ref{susol1}) into eq. (\ref{bi0}) taking
into account eq. (\ref{susol2}) the result is%
\begin{equation}
1-\left(  \overset{.}{v}\right)  ^{2}\sim\frac{W(T\left(  z+v(t)\right)
)}{W(T\left(  z\right)  )}. \label{supsol9}%
\end{equation}
Strictly speaking, eq. (\ref{supsol9}) is, in general, not consistent unless
the right hand side turns out to be independent on $z$. However, in many
soliton-like solutions, for large values of the argument the soliton field
($T$ in this case) rapidly approaches a constant value: in this approximation
eq. (\ref{supsol9}) provides with a detailed description of the evolution;
when $W=0$, the Born-Infeld case is recovered. As an example, let us consider
the case%
\[
W(T)\sim T;
\]
from eq. (\ref{supsol9}) it follows%
\begin{equation}
1-\left(  \overset{.}{v}\right)  ^{2}\sim\frac{T\left(  z+v(t)\right)
}{T\left(  z\right)  }. \label{supsol11}%
\end{equation}
Thus, if one considers a tachyon field which for large positive values of its
argument is exponentially decaying%
\[
T\left(  z\right)  \underset{z\rightarrow\infty}{\sim}\exp(-bz)\quad\ b>0,
\]
then one obtains from eq. (\ref{supsol11})%
\begin{align}
\int^{v}\frac{dv^{\prime}}{\sqrt{1-\exp(-bv^{\prime})}}  &  \sim
t\Rightarrow\nonumber\\
v  &  \sim t^{2},\ t\ small\label{supsol12}\\
v  &  \sim t,\ t\ large. \label{supsol13}%
\end{align}
As one would expect, when $v$ is large and $W(T)$ is highly suppressed one
gets a light-like wave solutions; rather, for small $v$ there are deviations
from the light-like behavior.

An interesting class of non trivial curved backgrounds on which the above
scheme also works is the following%
\begin{align}
g  &  =2d(z-t)d(z+t)+R^{2m}(z-t,x^{i},y^{a})\left(  \sum_{i=1}^{p}\left(
dx^{i}\right)  ^{2}\right) \nonumber\\
&  +R^{2n}(z-t,x^{i},y^{a})\left(  \sum_{a=1}^{q}\left(  dy^{a}\right)
^{2}\right)  ,\label{brametri}\\
0  &  =mp+nq.\nonumber
\end{align}
Indeed, metrics with the above block structure are ubiquitous in the low
energy description of p-Branes in supergravity. The metric (\ref{brametri}) is
of cosmological interest since, according to whether one considers the $y^{a}$
or the $x^{i}$ as ''extra-dimensions'' to be compactified, it can represent
(according to the matter fields accounting for the sources of the Einstein
equations) space-times in which the volume of the extra-dimensions is
shrinking while the ''macroscopic'' dimensions are expanding. As far as
tachyon fields depending only on $z$ and $t$ are concerned, on such curved
space-times eq. (\ref{eff1}) and eq. (\ref{nocau}) keep their flat space-time
form unchanged so that the above discussion also holds in these cases. In
particular, any exact solution of the flat purely electric Born-Infeld theory
depending only on $z$ and $t$ remains an exact solution on these back-grounds too.

\section{The implications of a sine-Gordon description}

In the previous section it has been shown that the splitting of eq. (\ref{a2})
into eq. (\ref{eff1}) and eq. (\ref{nocau}) can be useful: in particular, it
came out that eq. (\ref{eff1}) can be considered as the effective dynamical
equation while eq. (\ref{nocau}) singles out, among the solutions of eq.
(\ref{eff1}), the ones fulfilling natural boundary conditions. Thus, eq.
(\ref{eff1}) and the action (\ref{a1}) provide with a good description of
tachyon dynamics supporting interesting string-theoretic conjectures.

It is worth to recall here some generic features of the tachyon potential
$V(T)$\ which are commonly believed to be true: $V(T)$ should be an even
function of $T$, it should have a maximum for $T=0$, it should have vanishing
minima for $T\rightarrow\pm\infty$. Let us represent the tachyon potential as
follows
\begin{equation}
V(T)=\exp\left[  -P(T)\right]  ,\quad P(T)=P(-T),\quad P^{\prime}(0)=0,\quad
P^{\prime\prime}(0)>0, \label{pretac}%
\end{equation}
so that, at least for small $T$,
\begin{equation}
P(T)=a_{1}+a_{2}T^{2}+o(T^{4}),\ a_{2}>0. \label{bps4}%
\end{equation}
However it is more convenient to write $P(T)$ in a different way
\begin{equation}
P(T)=\frac{m^{2}}{\beta^{2}}\cos\left(  \beta T\right)  +o(T^{4})
\label{sinegordon1}%
\end{equation}
where $\beta$ is an adimensional coupling constant, $m$ is a mass parameter
and, to make unimportant the violation of the condition to have vanishing
minima for $V(T)$ when $T\rightarrow\pm\infty$, one has to assume $\beta\ll1$
in such a way that the first minima of $P(T)$ occurs for very large $T$. It is
worth to stress that there is no lack of generality in the parametrization
(\ref{sinegordon1}): provided $\beta\ll1$, the form (\ref{sinegordon1}) of the
tachyon potential encodes generic characteristics of $V(T)$ which are believed
to be true.

The advantage of this parametrization of $P(T)$ is, at least, twofold:
firstly, this allows to uptake the fact that the action (\ref{aziobps}) is
completely integrable both at classical and quantum level; more importantly,
the powerful theory of perturbations of integrable models could come into
play\footnote{To provide with a complete list of references on this theme is a
hopeless task: as far as the connection with tachyon dynamics is concerned,
the results in \cite{DMS96} could be very important in a future perspective.}.
It could be possible to perform a direct comparison between the well known
mass spectra and $S$-matrix of the sine-Gordon model and some very interesting
features of tachyon dynamics, such as the close relation between (the
tachyonic description of) non BPS and BPS Brane\footnote{In \cite{Se98} it has
been conjectured that a tachyonic soliton (kink) on the world-volume of an
unstable D(p+1)-brane is a BPS Dp-brane.} and the arising of perturbative
string states from D-Brane anti-D-Brane annihilation, which are believed to be
true but of which a fully satisfactory theoretical description is still
lacking\footnote{There are many papers supporting this modern understanding of
tachyon dynamics such as \cite{MSZ00} \cite{Da01}, an updated review with the
main recent achievements and a complete list of references is, for example,
\cite{He04}.}.

Let us recall some known features of sine-Gordon theory \cite{ZZ79}: it has a
discrete symmetry
\[
T\rightarrow T+\frac{2\pi}{\beta}%
\]
which is spontaneously broken at $\beta^{2}<8\pi$ which is the case of
interest in this context since the parametrization (\ref{sinegordon1}) can
provide with an effective description of tachyon dynamics only if $\beta\ll1$.
In this domain the theory is massive and its particles spectrum consists of
soliton-anti-soliton pair with equal masses and a certain number of quantum
breathers (the number $n_{B}$ of quantum breathers fulfills
\[
n_{B}=\left[  \frac{8\pi}{\beta^{2}}-1\right]
\]
where, in the above equation, the square brackets denote the integer part).
The (anti)solitons carry a (negative) positive topological charge
\begin{equation}
Q=\frac{\beta}{2\pi}\int_{-\infty}^{+\infty}\partial_{x}T(x,y)dx \label{sine2}%
\end{equation}
while the neutral breathers are soliton-anti-soliton bound states with
energies
\begin{equation}
E_{B}^{(n)}=2M_{s}\sin\left(  \frac{n\pi}{2(\frac{8\pi}{\beta^{2}}-1)}\right)
,\ n\leq n_{B},\ M_{s}=\frac{8m}{\beta^{2}}, \label{sine3}%
\end{equation}
where $M_{s}$ is the soliton mass.

The above spectrum is characteristic of sine-Gordon model but the main
features of the spectrum are shared by many field theoretical models admitting
solitonic solutions: the cross over from low energy states (interpreted as
soliton-anti-soliton bound states) to high energy states (solitons and
anti-solitons) is similar to the one in eq. (\ref{sine3}) in which, after a
finite number of low energy states (with energies of order $m$, the mass of
the ''\textit{phonons}'') there are solitons and anti-solitons (with energies
of order $\frac{8m}{\beta^{2}}$ where $\beta$ is the small coupling constant),
see, for example, \cite{VS00} and references therein.

To make contact with the action (\ref{a1}) the identification between the
string theoretic and the sine-Gordon parameters is the following:
\begin{equation}
m^{2}\rightarrow\alpha^{-1},\ \beta^{2}\rightarrow g_{S}=\left\langle \exp
\phi\right\rangle \label{sinepara}%
\end{equation}
$g_{S}$ being the string coupling constant. This identification can be
achieved for example, by recalling that tachyonic action is multiplied by
$\frac{1}{g_{S}}$ and, consequently, such a factor has to lie also in front of
the sine-Gordon action describing tachyon dynamics. Consequently, the
$\frac{1}{g_{S}}$ factor in front of the sine-Gordon action can be absorbed in
the Lagrangian as follows
\[
\frac{1}{g_{S}}(\nabla^{c}T\nabla_{c}T+W)\rightarrow\left(  \frac{\nabla^{c}%
}{g_{S}}\right)  \left(  g_{S}^{1/2}T\right)  \left(  \frac{\nabla_{c}}{g_{s}%
}\right)  \left(  g_{S}^{1/2}T\right)  +\left(  \frac{1}{g_{S}}W\right)  ,
\]
so that any $T$ is replaced by $\left(  g_{S}\right)  ^{1/2}T$ (which tells
that we are near the maximum of the tachyon potential at $T=0$), the
derivatives are rescaled by $1/g_{S}$ (which tells that the derivatives of the
tachyon are large when the tachyon is displaced from the local maximum) and
the constant $1/g_{S}$ appears in front of the self-interaction potential.

\subsection{Physical consequences of the IM description}

Whatever the details of the underlying unstable Brane are, from eqs.
(\ref{sine3}) and (\ref{sinepara}) the mass of the tachyonic soliton is of
order of a D-Brane tension
\[
M_{s}\sim\frac{1}{g_{S}}%
\]
and this is in a perfect agreement with \cite{Se98}. Also intriguing is the
interpretation of the breathers\ of sine-Gordon theory: they represent
soliton-anti-soliton bound states. According to \cite{Se98}, this allows to
think at breathers as D-Brane-anti-D-Brane bound states, and, from eqs.
(\ref{sine3}) and (\ref{sinepara}), the energies of such bound states are
\begin{equation}
\left(  E_{B}^{(n)}\right)  ^{2}\sim\frac{\alpha^{-1}}{\left(  g_{S}\right)
^{2}}\left(  \sin\left(  \frac{n\pi}{2(\frac{8\pi}{g_{S}}-1)}\right)  \right)
^{2}. \label{pertregime}%
\end{equation}
When the string coupling constant is very small (and this happens precisely
when this description does work: $\beta\ll1$), deeply into the perturbative
regime, the energies of the above (non perturbative) bound states of D-Brane
and anti-D-Brane turn out to be
\[
\left(  E_{B}^{(n)}\right)  ^{2}\sim c_{0}n^{2}\alpha^{-1}%
\]
which are nothing but the (square of the) energies of integer numbers of
fundamental (perturbative) strings. This provides the ideas in \cite{BHY00}
\cite{GHY00} (that, under certain circumstances, when a D-Brane and an
anti-D-Brane annihilate perturbative string states can emerge) with a sound,
fully dynamical, support. The above considerations encourage to further pursue
this point of view.

What happens if we add to $P(T)$ another piece which takes into account that a
sine-Gordon description should be effective only for small value of the
tachyon field? It has been shown in \cite{DMS96} that a generic perturbation,
removing the degeneracy among the different classical vacua, lead to the
confinement of (topologically) charged solitons (which, in this picture,
represent charged D-Branes, the non perturbative stable states of string
theory) while the breathers\ (which can be identified with perturbative string
states) are left in the physical spectrum\footnote{In \cite{DMS96} it has also
been shown that, if the perturbation does not remove completely the degeneracy
among the different classical vacua, some solitons survives in the physical
spectrum. The authors of \cite{DMS96} analyzed in details a particular
perturbation of sine-Gordon theory but their qualitative conclusions about the
confinement are general.} thus providing with a clear physical description of
the not yet fully understood interplay between perturbative and non
perturbative regime which underlies several crucial issues such as tachyon
condensation (for a detailed updated review see \cite{He04}) and whether and
how confinement occurs in the weakly coupled limit of string theory (see
\cite{Se98} \cite{BHY00} \cite{GHY00} and references therein).

The main lesson is that a description in term of (perturbation of) integrable
models of tachyon dynamics could capture non perturbative phenomena: it
naturally encompasses the seeming dichotomy between high and low energy
phenomena. It could be fruitful to repeat the analysis in \cite{Se02I} using
Boundary Integrable Field Theory instead of Boundary Conformal Field Theory.

\section{Conclusion}

A suitable splitting of the tachyon field equation into two equations, the
first one representing an effective dynamical equation and the second one a
sort of constraint related to (the absence of) caustics, is helpful to find
exact solutions; the scheme also works on curved space-times of cosmological
interest. It has been derived in a simple way the (already known) property
that \textit{square-integrable} wave-like solutions of Maxwell theory are
solutions of Born-Infeld theory too. In fact, this is not true if one
considers \textit{non} \textit{square-integrable} wave-like solutions (which,
when asymptotically constant, have $\delta-$like singularities somewhere) of
Maxwell theory: this is related to the fact that Born-Infeld theory has the
property that the electric field is bounded from the above. A similar
phenomenon has also been displayed in the standard Polyakov action for strings
on flat space-times: this is related to the fact that Polyakov action can be
derived from the Nambu-Goto action which is of Born-Infeld type. It has also
been argued that a splitting of the (logaritm of the) tachyon potential into
an integrable sine-Gordon term plus a perturbation can have highly non trivial
consequences. The spectrum of sine-Gordon theory provide with a very sound
support to two related conjectures: the first asserts that a tachyonic soliton
(kink) on the world-volume of an unstable D(p+1)-Brane is a BPS Dp-Brane; the
second contends that, under certain circumstances, when a D-Brane and an
anti-D-Brane annihilate perturbative string states can emerge. If one perturbs
the sine-Gordon term with terms which remove the degeneracy among the
different classical vacua, then the D-Branes get confined and one is only left
with the perturbative string spectrum.


\bigskip

\begin{thebibliography}{99}
\bibitem{AL04}J.M. Aguirregabiria, R. Lazkoz, \textit{Phys.Rev.} \textbf{D}69
(2004) 123502; M. C. Bento, O. Bertolami, A. Sen, \textit{Phys.Rev}.
\textbf{D}67 (2003) 063511; J.S.Bagla, H.K.Jassal, T.Padmanabhan,
\textit{Phys.Rev.} \textbf{D}67 (2003) 063504; J. M. Cline, H. Firouzjahi, P.
Martineau, \textbf{JHEP} 0211 (2002) 041; T.Padmanabhan, \textit{Phys.Rev}.
\textbf{D}66 (2002) 021301; Y. Piao, R. Cai, X. Zhang, Y. Zhang,
\textit{Phys.Rev}. \textbf{D}66 (2002) 121301; M. Sami,
\textit{Mod.Phys.Lett.} \textbf{A}18 (2003) 691; G. Shiu, S. H. Tye, I.
Wasserman, \textit{Phys.Rev}. \textbf{D}67 (2003) 083517.

\bibitem {BR00}E.A. Bergshoeff, M. de Roo, T.C. de Wit, E. Eyras, S. Panda,
\textbf{JHEP} 0005 (2000) 009.

\bibitem {BH93II}M. Bordemann, J. Hoppe, \textit{Phys.Lett}. \textbf{B}317
(1993) 315, \textit{Phys.Lett}. \textbf{B}325 (1994) 359.

\bibitem {BHY00}O. Bergman, K. Hori, P. Yi, \textit{Nucl.Phys}. \textbf{B}580
(2000) 289; P. Yi,\textit{ Nucl.Phys}. \textbf{B}550 (1999) 214.

\bibitem {Cho99}K. Choi, \textit{Phys.Rev}. \textbf{D}62 (2000) 043509.

\bibitem {Da01}S. Dasgupta, T. Dasgupta, \textbf{JHEP} 0106 (2001) 007.

\bibitem {DMS96}G. Delfino, G. Mussardo, P. Simonetti, \textit{Nucl.Phys}.
\textbf{B}473 (1996) 469; G. Delfino, G. Mussardo, \textit{Nucl.Phys}.
\textbf{B}516 (1998) 675.

\bibitem {FK02}G. Felder, L. Kofman, A. Starobinsky \textbf{JHEP} 0209 (2002)
026; G. N. Felder, L. Kofman, \textit{Phys.Rev}. \textbf{D}70 (2004) 046004.

\bibitem {Ga00}M. R. Garousi, \textit{Nucl.Phys}. \textbf{B}584 (2000) 284.

\bibitem {GST04}M. R. Garousi, M. Sami, S. Tsujikawa, \textit{Phys.Rev}.
\textbf{D}70 (2004) 043536

\bibitem {Gi97}G. W. Gibbons,\textit{ Nucl.Phys.} \textbf{B}514 (1998) 603.

\bibitem {GHY00}G. W. Gibbons, K. Hori, P. Yi, \textit{Nucl.Phys.}
\textbf{B}596 (2001) 136.

\bibitem {He04}M. Headrick, S. Minwalla, T. Takayanagi,
\textit{Class.Quant.Grav}. \textbf{21} (2004) S1539.

\bibitem {Ho94}J. Hoppe, hep-th/9402112, hep-th/9503069.

\bibitem {KKLT03}S. Kachru, R. Kallosh, A. Linde, S. P. Trivedi,
\textit{Phys.Rev.} \textbf{D}68 (2003) 046005.

\bibitem {Klu00}J. Kluson, \textit{Phys.Rev.} \textbf{D}62 (2000) 126003.

\bibitem {KL02}L. Kofman, A. Linde, \textbf{JHEP} 0207 (2002) 004.

\bibitem {KN03}D. Kutasov, V. Niarchos, \textit{Nucl.Phys}. \textbf{B}666
(2003) 56.

\bibitem {MSZ00}N. Moeller, A. Sen, B. Zwiebach, \textbf{JHEP} 0008 (2000) 039.

\bibitem {Mu02}S. Mukohyama,\textit{ Phys.Rev}. \textbf{D}66 (2002) 123512; A.
Sen, \textit{Phys.Rev.} \textbf{D}68 (2003) 066008.

\bibitem {Se98}A. Sen, \textbf{JHEP} 9812 (1998) 021.

\bibitem {Se02I}A. Sen, \textbf{JHEP} 0204 (2002) 048, \textbf{JHEP} 0207
(2002) 065, \textbf{JHEP} 0210 (2002) 003. \ \ \ \ \ \ \ \ \ \ \ \ \ \ \ \ \ \ \ \ \ \ \ \ \ \ \ \ \ \ \ \ \ \ \ \ \ \ \ \ \ \ \ \ \ 

\bibitem {VS00}A. Vilenkin, E. P. S. Shellard, ''\textit{Cosmic Strings and
Other Topological Defects}'' (Cambridge University Press, 2000).

\bibitem {ZZ79}A. B. Zamolodchikov, Al. B. Zamolodchikov, \textit{Ann. Phys}.
\textbf{120} (1979), 253 and references therein.
\end{thebibliography}
\end{document}